\documentclass[final,5p,times]{elsarticle}

\usepackage[linktocpage=true,unicode=true,ps2pdf]{hyperref}
\usepackage{hypcap}
\usepackage{amsmath,amssymb,amsfonts}
\usepackage{graphicx}
\usepackage{color}

\graphicspath{{./figs/}}
\begin{document}

\begin{frontmatter}

\title{Coherent and random UHECR  Spectroscopy of Lightest Nuclei  along $Cen_A$:  \\
  Shadows on GZK Tau Neutrinos  spread in a near sky and time}
\author{Daniele Fargion}
\ead{daniele.fargion@roma1.infn.it}
\address{Dipartimento di Fisica Teorica, Universit\`{a} di Roma 1, Sapienza
and INFIN --Pl. A. Moro 2, 00185, Rome, Italy.}

\begin{abstract}
As we did noticed since earliest  UHECR anisotropy findings \cite{Auger-Nov07}, today ICRC09 AUGER updated maps and clustering confirms our understanding of CenA as the main nearby UHECR source, whose events are mostly lightest nuclei, as He. The events are spread by galactic fields.  The  main UHECR events along CenA are spreading vertically (respect to horizontal spiral galactic fields) by random (and  a final coherent) Lorentz forces.  The He-nuclei dominance is also well probed  by detailed AUGER composition data. The Lightest nuclei are  still compatible with most recent Hires composition results. Such lightest  UHECR nuclei are  opaque even to nearest Universe. Offering a very narrow local (ten Mpc) view. Explaining at best the otherwise puzzling Virgo persistent absence.  The He disruption should lead to halves energy tail clustering (along the same $Cen_A$ group), events of deuterium, $ He^3$ and protons (or unstable neutrons). They may rise in  smeared cluster of events along the same $Cen_A$ group edges at energies among $1.5-4.5 \cdot 10^{19}$ eV. The  He-like UHECR photo-disruption  should produce also energetic gamma (extremely opaque with CMB and therefore local and hard to detect in nosy knee CR) and GZK neutrinos  at tens-hundred PeVs energy.  Contrary to widely expected GZK $\nu$ at EeVs energy  (due to still a widely believed, by others, UHECR proton  nature). Our foreseen UHE  neutrino by UHECR Lightest nuclei, at tens-a hundred PeV, may rise in AUGER and TA telescopes via  quite inclined (about $5^o$ ) upward fluorescent Tau air-showers  at nearest ($1-4$ km)  horizons, at middle-low telescope view zenith angles (below $20^o$). Their air-showers will be  opening and extending in widest azimuth angle (above $120^o$) and  fastest angular velocity spread. Contrary to far EeV tau expected by popular  GZK neutrinos. Even hard to observe these   Tau Air-showers at  tens and hundred PeVs may soon shine (in a peculiar way)  at near telescope views and within a short (years) time, both in AUGER, HEAT and TA.  \\
\end{abstract}


\end{frontmatter}


\section*{Introduction}
The understanding of UHECR is generally related to their spectra and (eventual) GZK cut off \cite{Greisen:1966jv},\cite{za66}. However a
complete understanding requires both spectra and map. At best the energy map identity of each UHECR event. The earliest UHECR map, from historic one of  Fly Eyes, to last decade AGASA, the more recent HIREs and the final AUGER ones, led to excitement, contradictions and surprises. Since November 2007 \cite{Auger-Nov07}, the GZK steepening in UHECR (following HIRES \cite{Hires06}) and the earliest AUGER anisotropy seemed to firmly  point to a GZK cutoff, by a map overlapping the Super-Galactic Plane (SGP), favoring a consequent proton composition \cite{Auger-Nov07}. However the same AUGER data on composition (ICRC 2007) hinted a heavy (or light) nuclei contrary to nucleon. And Virgo sources were missing in earliest AUGER map. Because of it I considered lightest nuclei : lightest UHECR cannot fly too far (above few or 10 Mpc) explaining the mysterious Virgo absence \cite{Fargion2008}\cite{Fargion2009}. I also suggested a key role of CenA as the main UHECR  nearby source: its event spread (by  these lightest nuclei) events overlap, by (perverse) chance, the SGP \cite{Fargion2008}, \cite{Fargion2009}, explaining the confusion. The very recent (preliminary) UHECR map (shown by AUGER collaboration  and published at HEP 2009 talk meeting) \cite{Hep09}, reconstructed on different background here, is including $58$ events recorded up to March 2009 at energies above nearly 55 EeV. We found   still Cen A as the major clustering source as well as the remarkable Virgo absence as the main anisotropy signal.  This map confirms three AUGER great results. The first is the endurance of the Virgo absence, over a huge solid angle area (within $10\%$ of the sky map, or $20\%$ of the visible AUGER  map), noted soon by few authors \cite{Gorbunov09},\cite{Fargion2008}, based on the same first AUGER map (2007). The second is the appearance  of new more spread isotropic events smearing (or masking) most of the apparent anisotropy along the Super-Galactic Plane. This imply the absence of the SGP signature and a minor  proton role. New GZK map seem to be considered now and well be discussed elsewhere. The third is the growing clustering of events along the Cen A AGN source, confirming a main remarkable anisotropy of AUGER sky. These events are spread by peculiar signature. Almost vertical to galactic plane. Indeed this nearest and most powerful source cluster in a narrow area, almost two percent of the sky, (or four percent of the AUGER sky) contains an over-abundance of nearly a quarter ($22\%$) of the whole event map. There is also inside the group a  thinner area, a string of event clustering along Cen A that is suggesting a random vertical  spectroscopy (by planar spiral galactic fields) of the CenA UHECR events. They are spread by energy and possibly by a few lightest nuclei charges orthogonally to Galactic fields. Because of it the consequent GZK cut and nuclei photodisintegration the UHECR will fragment into halves secondaries and they are showering in tens PeV gamma and neutrinos, spread into clustered string of events. Tau Neutrinos at tens PeV may skim the Earth, produce a tau, whose escape and decay at horizons is leading to tau air-showers. At this tens-hundred PeV energy tau will be detectable by fluorescence air-showering only within a  very near distance of AUGER (HEAT,TA) telescopes. In a very peculiar way. In near future (two-four years).
\section*{Coherent versus random spectroscopy}
There are two main spectroscopy of UHECR along galactic plane:
 A late nearby (almost local) bending by a nearest coherent galactic arm field, and a random one along the whole plane.
The coherent Lorentz angle bending $\delta_{Coh} $ of a proton UHECR (above GZK) within a galactic magnetic field  in a final nearby coherent length  of $l_c = 1\cdot kpc$ is
\begin{equation}
\delta_{Coh-p} \simeq
{2.3^\circ}\cdot \frac{Z}{Z_{H}} \cdot (\frac{6\cdot10^{19}eV}{E_{CR}})(\frac{B}{3\cdot \mu G}){\frac{l_c}{kpc}}
\end{equation}
The corresponding coherent  bending of an Helium UHECR at same energy, within a galactic magnetic field
  in a wider nearby coherent length  of $l_c = 2\cdot  kpc$ is
\begin{equation}
\delta_{Coh-He} \simeq
{9.2^\circ}\cdot \frac{Z}{Z_{He}} \cdot (\frac{6\cdot10^{19}eV}{E_{CR}})(\frac{B}{3\cdot \mu G}){\frac{l_c}{2 kpc}}
\end{equation}
The heavier of lightest nuclei that may be bounded from Virgo, Be, is bent by
$
\delta_{Coh-Be} \simeq
{18.4^\circ}\cdot \frac{Z}{Z_{Be}} \cdot (\frac{6\cdot10^{19}eV}{E_{CR}})(\frac{B}{3\cdot \mu G}){\frac{l_c}{2 kpc}}
$.  It should be noticed that the galactic magnetic field nearby and along $Cen_A$ direction is described by spiral galactic fields along the Milky Way Plane and (toward Cen A) as a plume (vertical to galactic plane). Both fields act respectively to  bend  the charges either coherently and-or randomly. We believe that the vertical incoherent spread (as estimated below) rules the bending for extragalactic  CenA. However nearest sources as eventual galactic ones may suffer a coherent bending. Such a coherent bending suffer an energy (inverse power law) dependence and a charge (linear) proportion as shown in equation above.
\subsection*{Random spectroscopy orthogonal to Galactic Plane}
The incoherent random angle bending a $\delta_{rm} $ while crossing a Galactic diameter of $ L\simeq{20 kpc}$ within a characteristic
 coherent length  $ l_c \simeq{2 kpc}$ is
\begin{equation}
\delta_{rm-p} \simeq
{8^\circ}\cdot \frac{Z}{Z_{H}} \cdot (\frac{6\cdot10^{19}eV}{E_{CR}})(\frac{B}{3\cdot \mu G})\sqrt{\frac{L}{20 kpc}}
\sqrt{\frac{l_c}{2 kpc}}
\end{equation}
\begin{equation}
\delta_{rm-He} \simeq
{16^\circ}\cdot \frac{Z}{Z_{He^2}} \cdot (\frac{6\cdot10^{19}eV}{E_{CR}})(\frac{B}{3\cdot \mu G})\sqrt{\frac{L}{20 kpc}}
\sqrt{\frac{l_c}{2 kpc}}
\end{equation}
The heavier  (but still lightest nuclei) bounded from Virgo are Li and Be:
$\delta_{rm-Li} \simeq
{24^\circ}\cdot \frac{Z}{Z_{Li^3}} \cdot (\frac{6\cdot10^{19}eV}{E_{CR}})(\frac{B}{3\cdot \mu G})\sqrt{\frac{L}{20 kpc}}
\sqrt{\frac{l_c}{2 kpc}}
$, $
\delta_{rm-Be} \simeq
{32^\circ}\cdot \frac{Z}{Z_{Be^4}} \cdot (\frac{6\cdot10^{19}eV}{E_{CR}})(\frac{B}{3\cdot \mu G})\sqrt{\frac{L}{20 kpc}}
\sqrt{\frac{l_c}{2 kpc}}
$.  It should be noted that the present anisotropy above GZK energy $5.5 \cdot 10^{19} eV$ might leave a tail of signals: indeed the photo disruption of He into deuterium, Tritium, $He^3$ and protons (and unstable neutrons), might rise as clustered events at half or quater of the energy.
  It is important to look for correlated tails of events, possibly in correlated strings at low $\simeq 1.5-3 \cdot 10^{19} eV$ along the $Cen_A$ train of events. In conclusion He like UHECR  maybe bent by a characteristic as large as  $\delta_{rm-He} + \delta_{Coh-He} \simeq 25.2^\circ$. Well within the observed CenA main event spread.
\section*{Maps  Composition of UHECR}
\subsection*{Why not protons, not iron, not C,N,O?}
  UHECR at energies up $1-2\cdot 10^{19}$ eV are ruled by protons as
  most mass composition show as well as their bending and homogeneity in sky are clearly probing.
  The absence of structure is simply related to the wide Lorentz bending by galactic and extragalactic fields.
  Their huge allowed cosmic volumes guarantee an isotropic map. Above $5.5 \cdot10^{19}$ eV their presence had to reflect a limited GZK volumes
  as the SGP and Local Universe, (that is now really unobserved). \emph{{In particular Virgo cluster, well beyond GZK cut off, had to rise as a main pointing clustered  bump in the UHECR } Sky.}  Moreover the presence of a wide smeared area of  events along Cen A (up to $20-30^o$) cannot be related to a more collimated proton bullet ( $4.6-8^o$). In conclusion:  No Virgo, no protons. For iron, as above the mass composition up few ten EeV stand for nucleon (and not heavy nuclei). However above  $4 \cdot 10^{19}$ eV AUGER favored heavy (or light) nuclei. Indeed the heaviest event, one or two, might be iron. So why not all iron? Once again because of  the absence of Virgo that avoid this possibility.
Indeed the iron $Z=26$ at $5.5 \cdot10^{19}$ eV will be bent up at least to $60^{o}$ by coherent fields up to $145^{o}$ by random fields. Their homogeneous spread  had to  fill the Virgo absent area. Moreover the observed clustering around CenA is too collimated to be indebt to iron. In conclusion once again: No Virgo, no iron.{Why not C,N,O ? Indeed light nuclei as C,N,O has been one of our choices \emph{if soon or later Virgo UHECR had  to rise}. However  Virgo  persistent absence   forced us to reject even this possibility.
\subsection*{Why He, Li,Be and D?}
Because of the  slant depth composition, because of the Virgo absence and because  the main spread angle from Cen A, it is obvious that the lightest nuclei, mostly Helium, are the very ideal candidate \cite{Fargion2008}, \cite{Fargion2009}  able to solve at once the main observed spectroscopy around CenA. Their consequent fragments ($He^3, D$) will trace by lower energy UHECR tail their presence; at tens-hundred PeV UHE photons secondaaies may hardly survive the GZK, but  neutrinos will rise , leading to a different rate and signature detection via Tau Air-showers \cite{Fargion2008},\cite{Fargion2009},\cite{Fargion2009b} .

\begin{figure*}[t]
\centering
\resizebox{0.6\hsize}{!}{\includegraphics[width=100pt, angle=0]{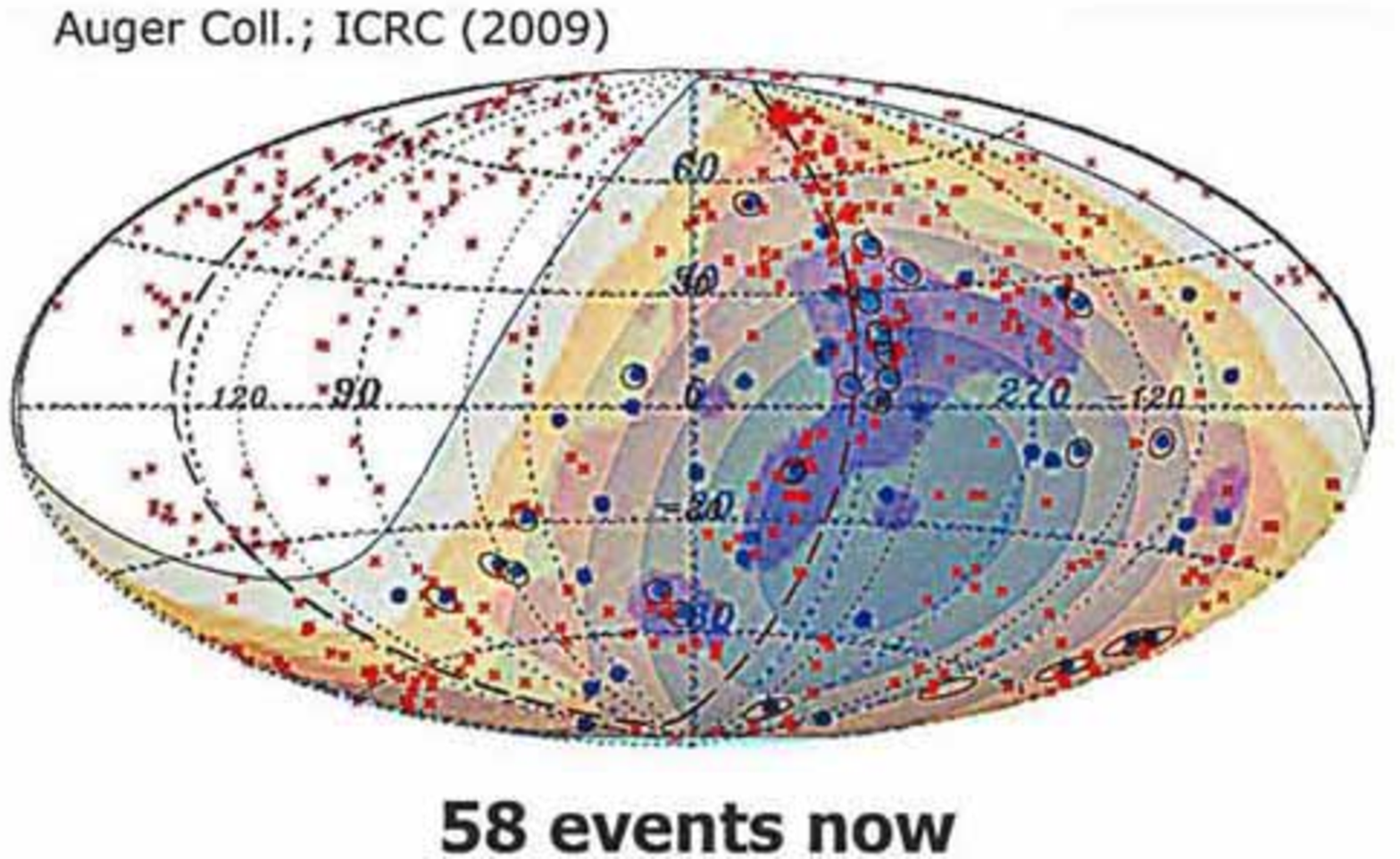}
\includegraphics[width=100pt, angle=0]{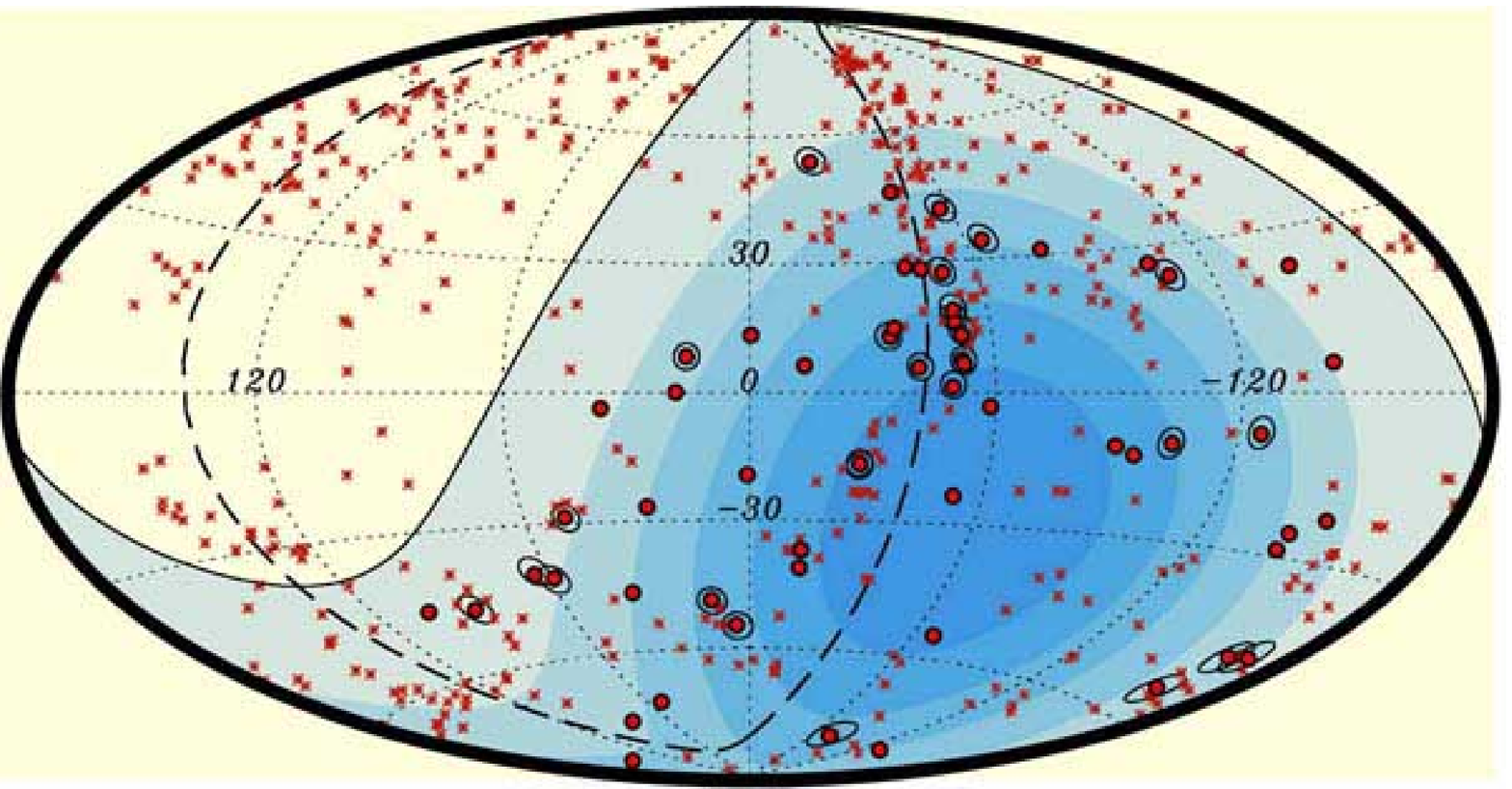}}
\caption{\label{01}
On the left the overlap AUGER map (2007-2009) as in slide n.22 \cite{{Hep09}}. The old events are within a wider ring.
This AUGER preliminary  map 2009, is shown  over the SWIFT BAT shadowed areas, ruled by gamma activities, assumed by AUGER
as the new GZK underlay map of UHECR. Even this shadowed area is more correlated to late UHECR events because of the remarkable Virgo absence (and because consequent arguments discussed here) the apparent correlation is not convincing. On the right side the old and the new event maps: the  Super-galactic Plane is the usual dashed line. The UHECR events, based on a preliminary map, rebuild by us here for clarity, it must be taken with great care because the very blurred preliminary ICRC09 source map used on the left side.
}
\end{figure*}
\begin{figure*}[t]
\centering
\resizebox{0.6\hsize}{!}{\includegraphics[width=100pt, angle=0]{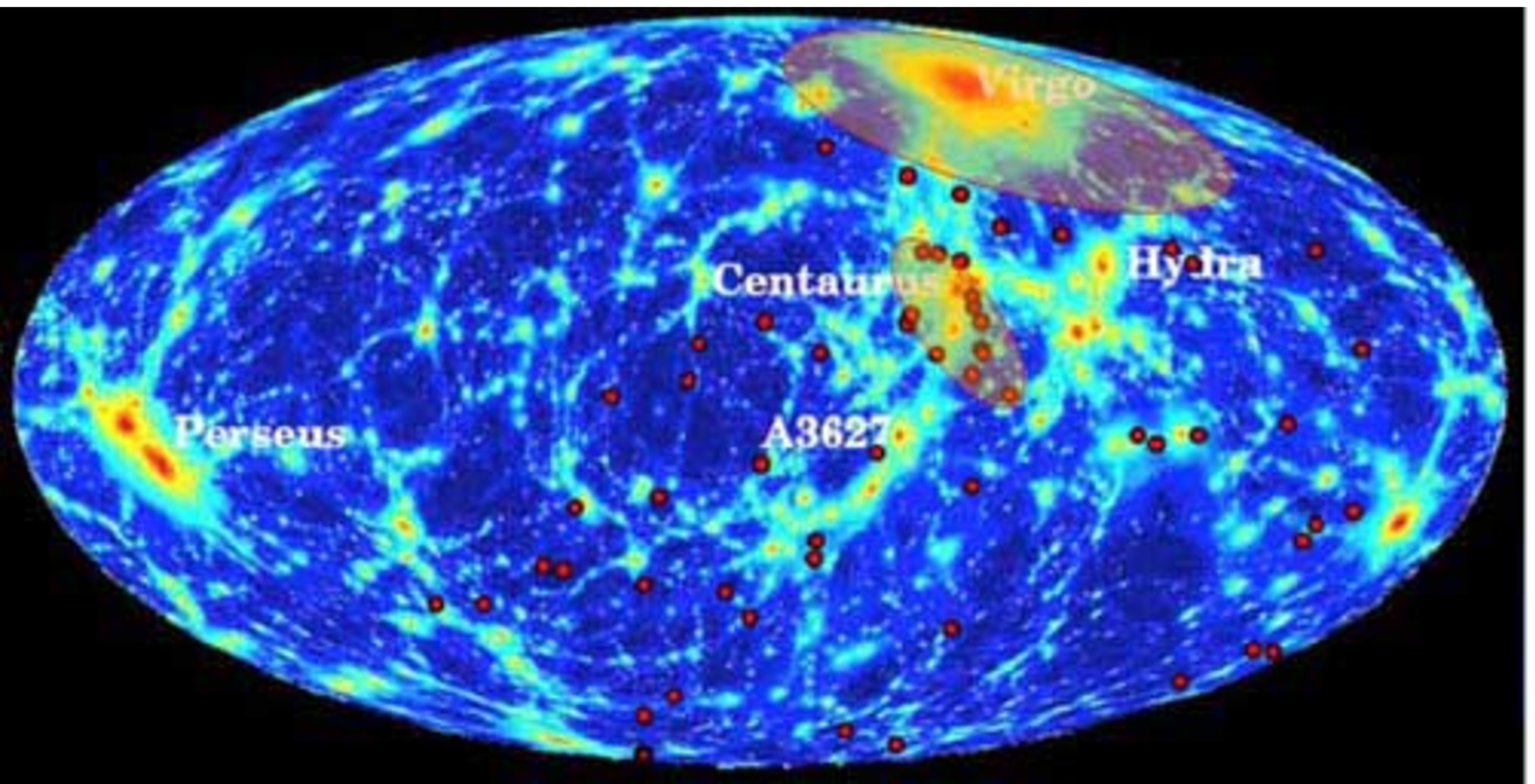}
\includegraphics[width=100pt, angle=0]{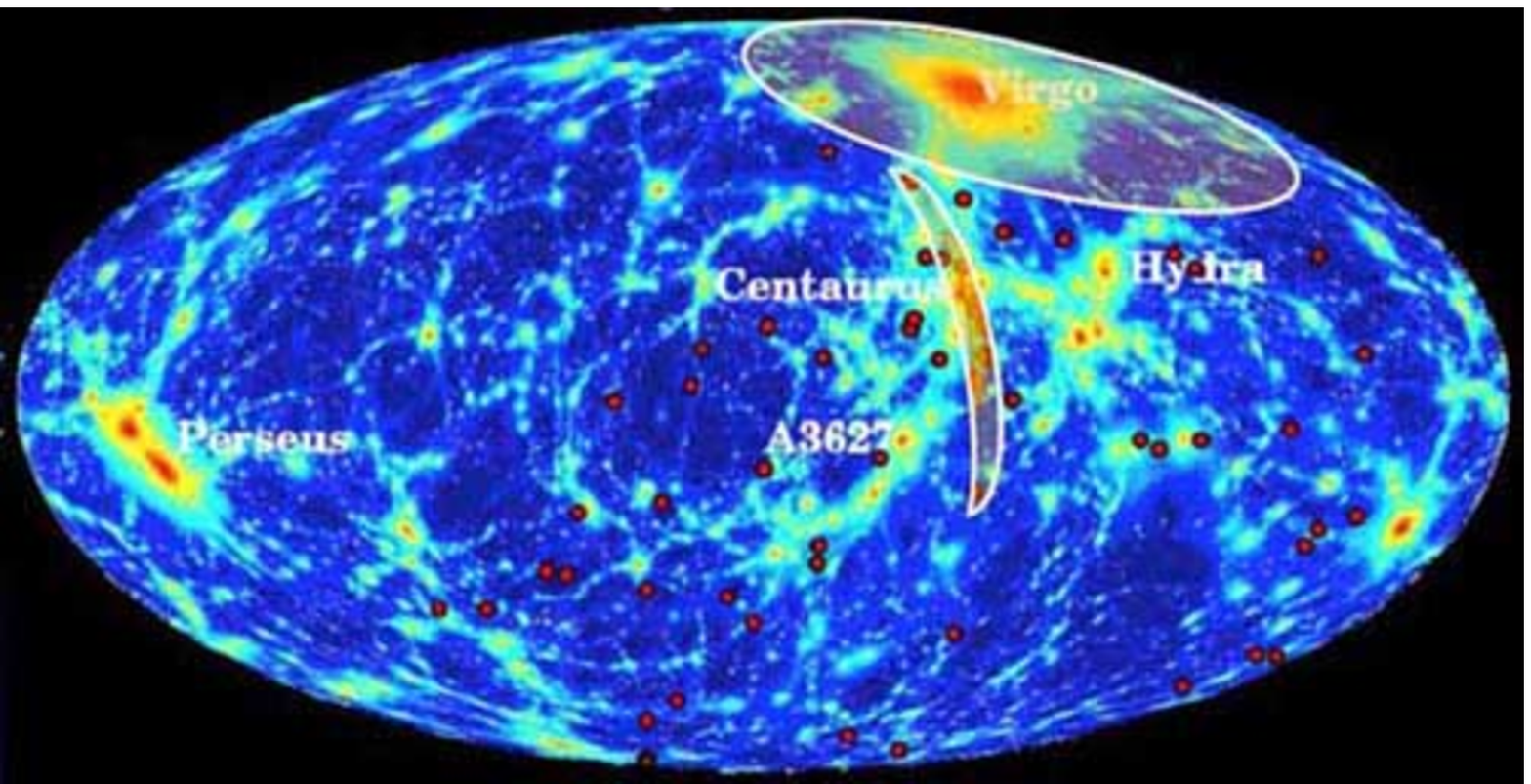}}
\caption{\label{02} The above AUGER UHECR events and an  underline cosmic Infrared map: the CenA  clustered event group are spread either in a very narrow oval (left), less than two (or four) percent of the sky or (right) into a thinner string (arc) area of events  (orthogonal to galactic plane)  . Both areas are small and crowded; they show remarkable  anisotropy. The left oval maybe due to both random (vertical)  and  a coherent (horizontal) bending due to a plume galactic field \cite{Fargion2009b}. The right narrower arc area  contains also the highest UHECR event recorded by Auger at its top; the wider arc area might contain  a tail of lower energy UHECR fragment events by D, $He^{3}$, at  few tens EeV, secondary of He or Be UHECR primary. Both  maps shows on the top the remarkable huge $empty$ Virgo area.
}

\end{figure*}
\begin{figure*}[t]
\centering
\resizebox{0.8\hsize}{!}{\includegraphics[width=100pt, angle=0]{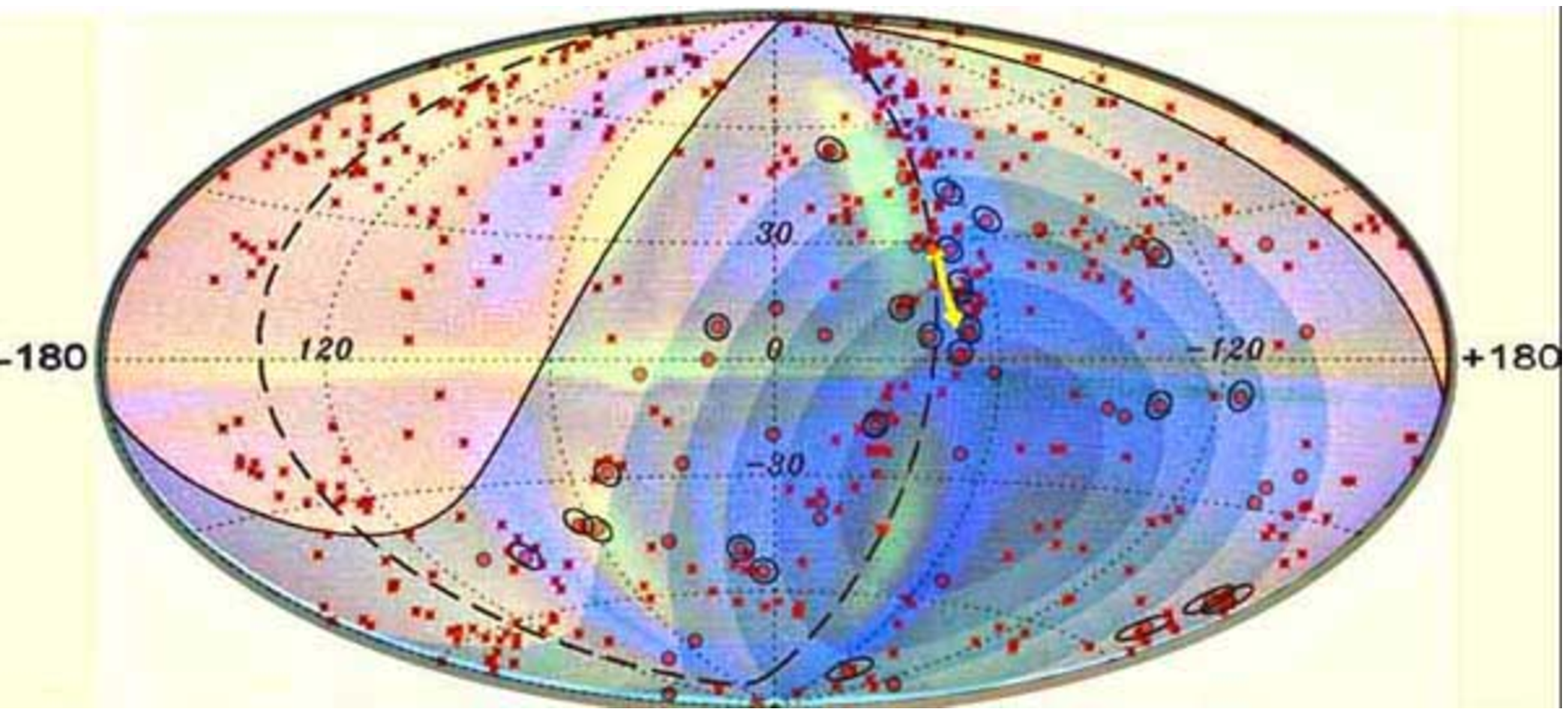}
\includegraphics[width=100pt, angle=0]{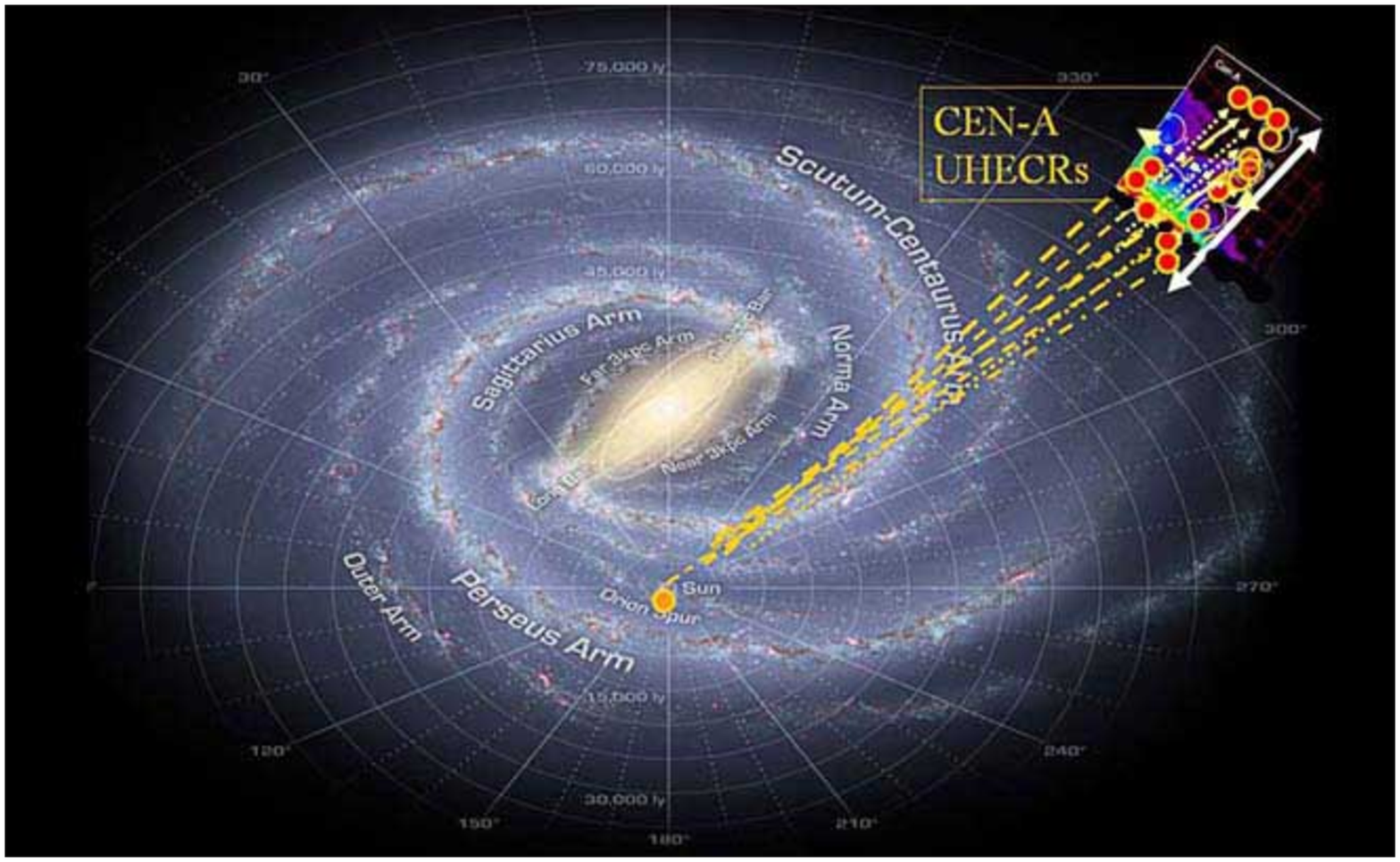}}
\caption{ \label{03} On the left the radio Ghz map with a consequent galactic (plane) and halo (twin plume vertical) magnetic fields and preliminary reconstructed AUGER 2009 events. Note the random vertical spread along the CenA  as marked by a white arrow. Because of the plume there is also an additional possible coherent deflection, explaining an asymmetric cluster on the yellow arrow left side.  On the right figure the schematic Galactic map in the oval of events may be spread by random (vertical white arrow) and by an eventual coherent (horizontal yellow dashed arrow) bending. In case of just a narrower arc clustering only random (white arrow) bending takes place \cite{Fargion2008},\cite{Fargion2009}.
}
\end{figure*}

\begin{figure*}[t]
\centering
\resizebox{0.8\hsize}{!}{\includegraphics[width=100pt, angle=0]{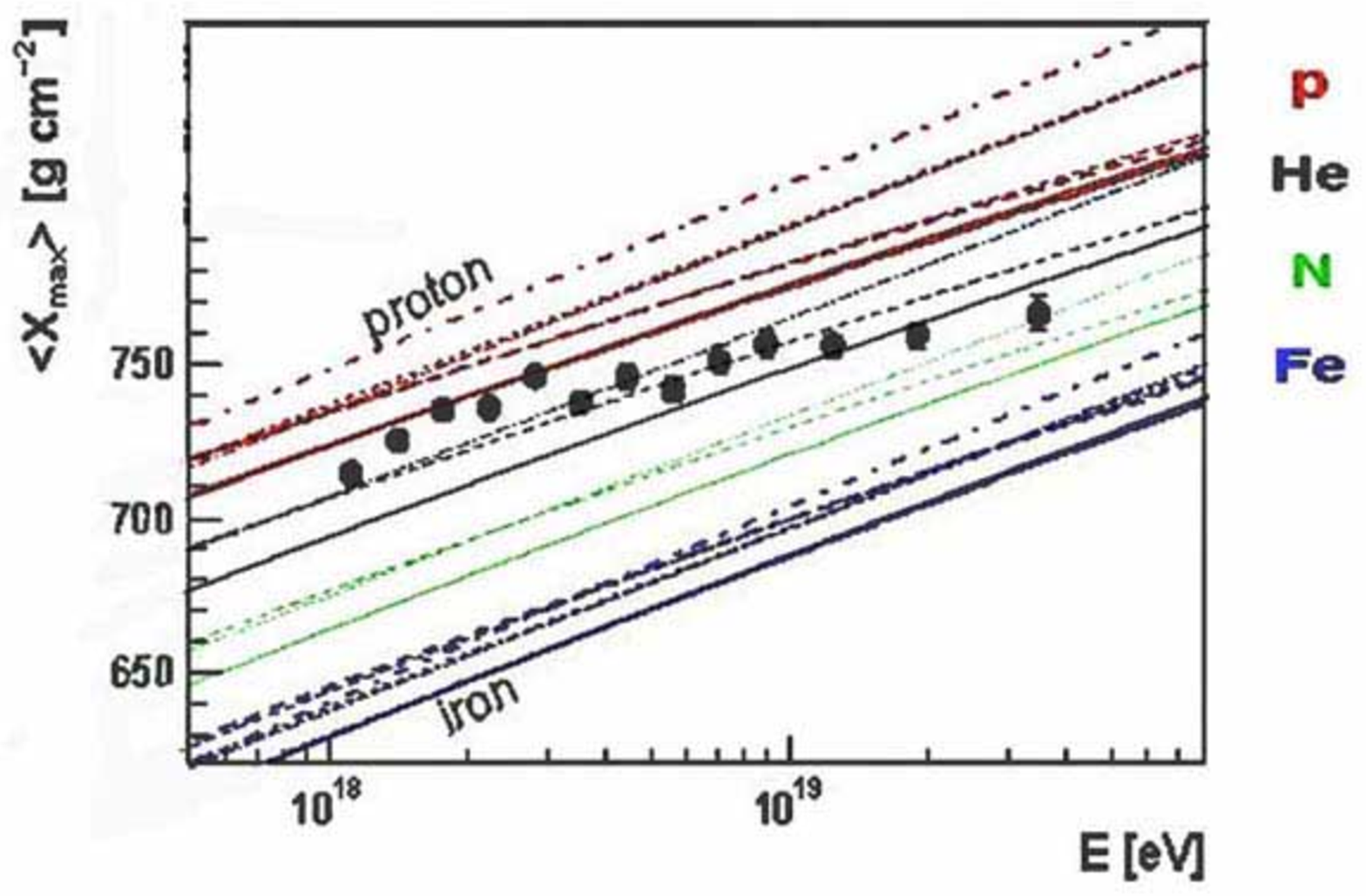}
\includegraphics[width=100pt, angle=0]{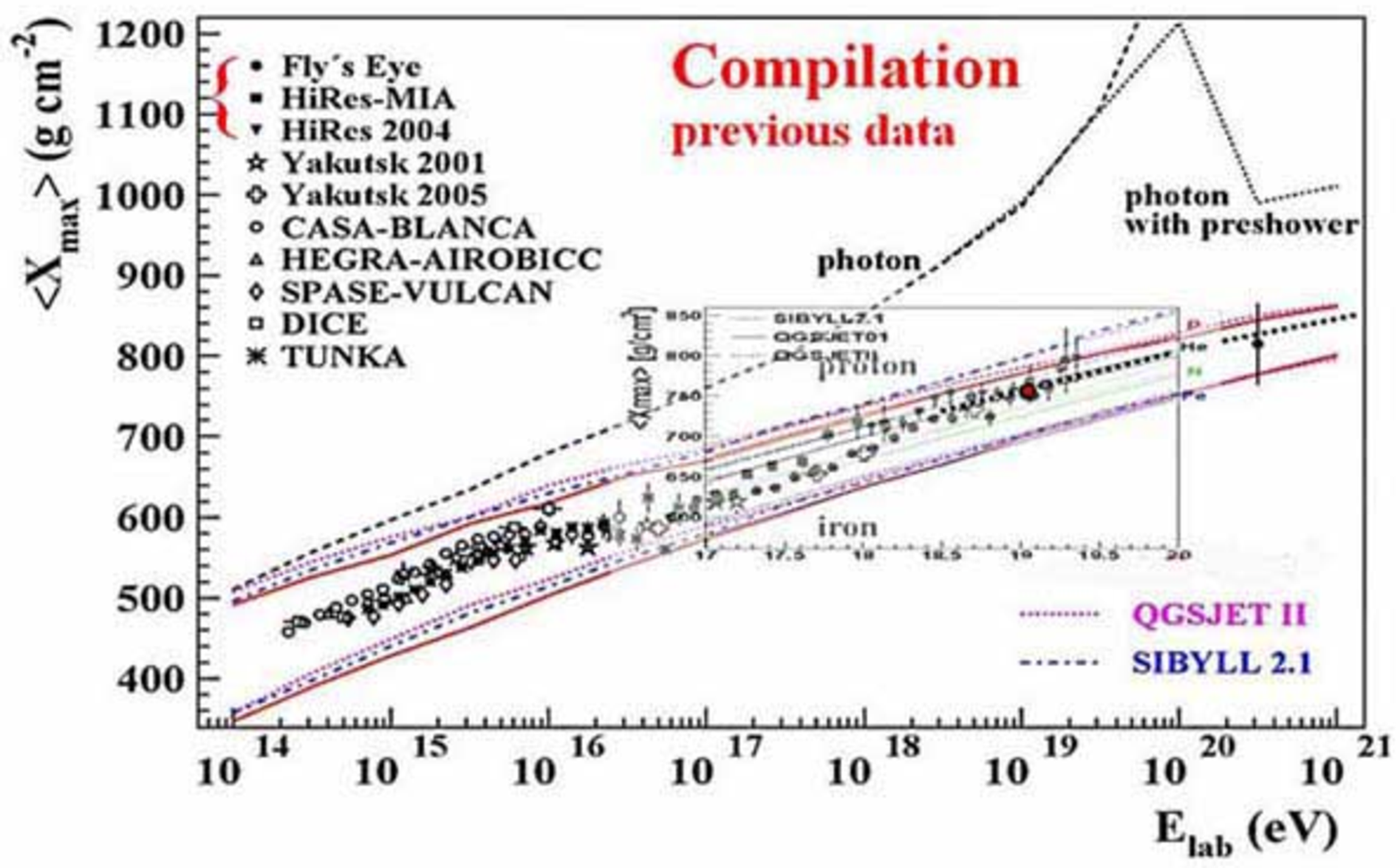}}
\caption{ \label{04} Left: The UHECR composition derived by different nuclear models and the last AUGER 2009 data shown in last meetings \cite{Hep09}. Right: The same for most energetic events within present (extended) nuclear models  \cite{Hep09}. The He nuclei candidature is at best fitting the observed air-shower composition shape. UHECR composition by HIRES is also compatible with same He nature of UHECR \cite{Fargion2008},\cite{Fargion2009}.
}
\end{figure*}

\begin{figure*}[t]
\centering
\resizebox{0.8\hsize}{!}{\includegraphics[width=100pt, angle=0]{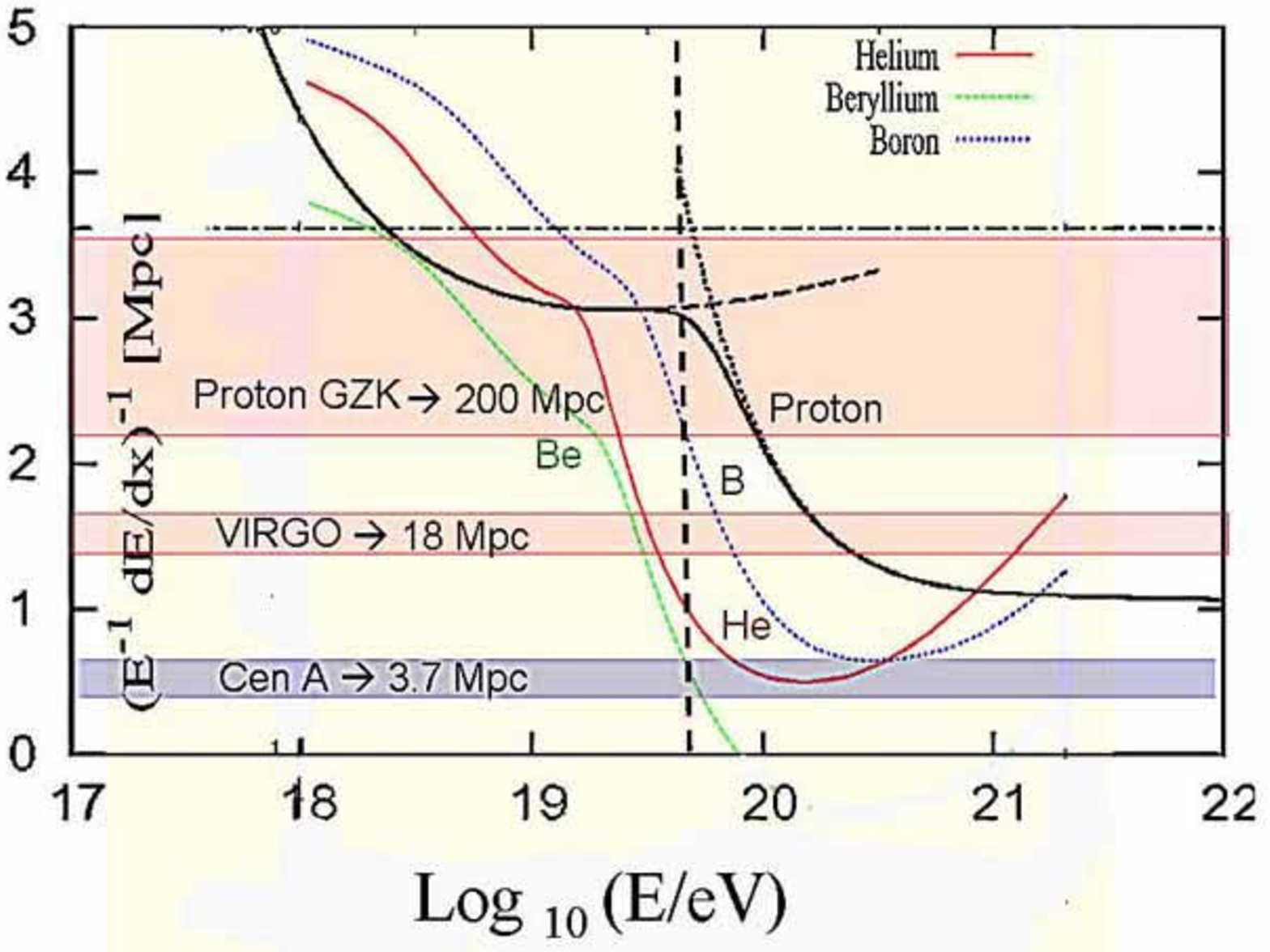}
\includegraphics[width=100pt, angle=0]{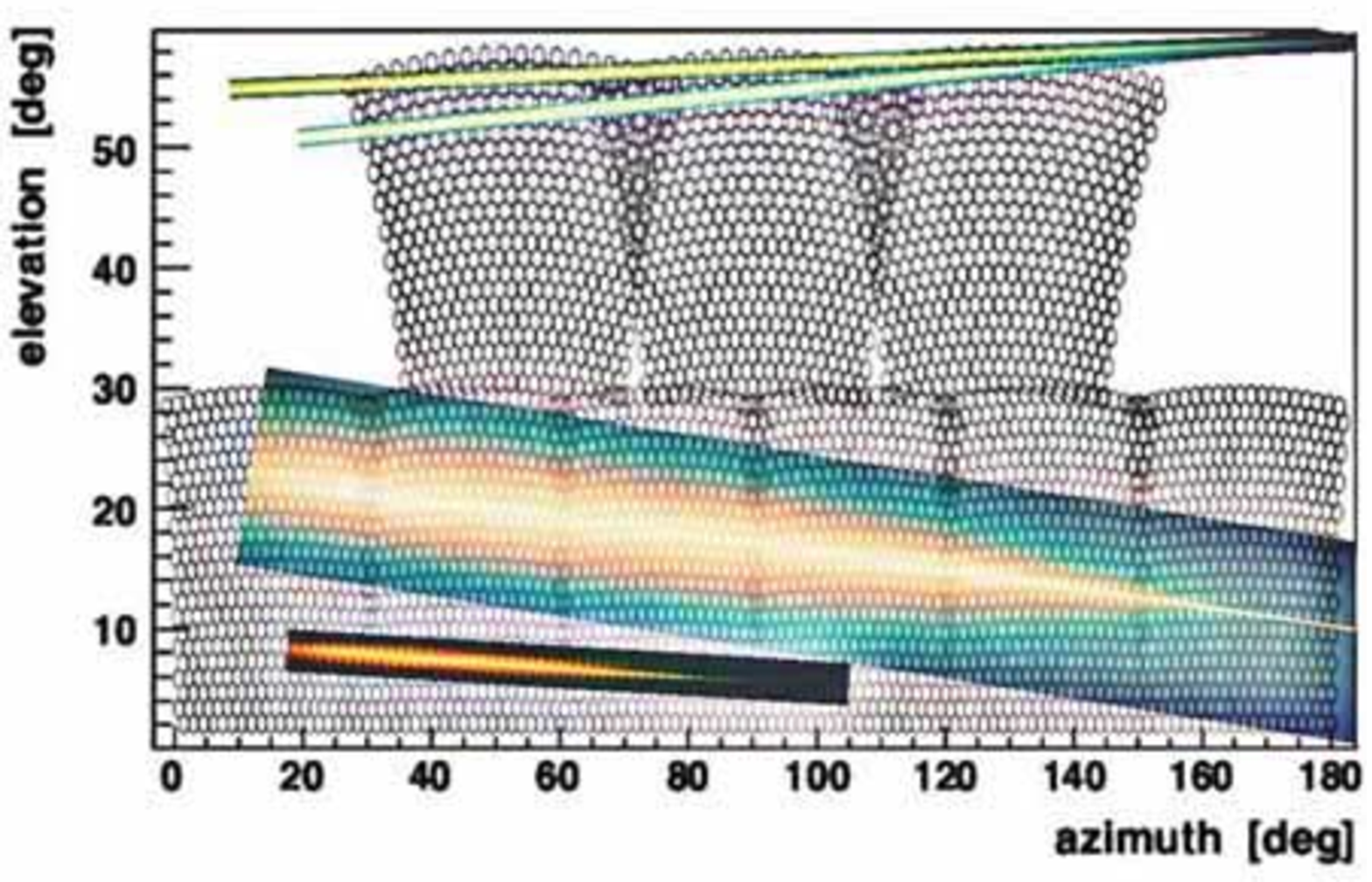}}
\caption{Left: The expected propagation distances for different lightest nuclei and proton UHECR. It is the short propagation distance for lightest nuclei to guarantee the Virgo absence. The energy threshold (55 EeV) may stop He,Be and partially B. Right: the expected GZK tau neutrinos and tau air-showers by lightest nuclei at tens PeV. They will appear as a widest ghost shadows very near the telescope, versus far EeV tau (for proton GZK) and far and high altitude horizontal hadron air-showers observable soon by HEAT telescopes possibly split by geomagnetic fields. AMIGA denser array may also reveal at ground SD detectors the Tens-hundred PeV tau airshowers \cite{Fargion2008},\cite{Fargion2009}.
}
\end{figure*}

\begin{figure*}[t]
\centering
\resizebox{0.8\hsize}{!}{\includegraphics[width=100pt, angle=0]{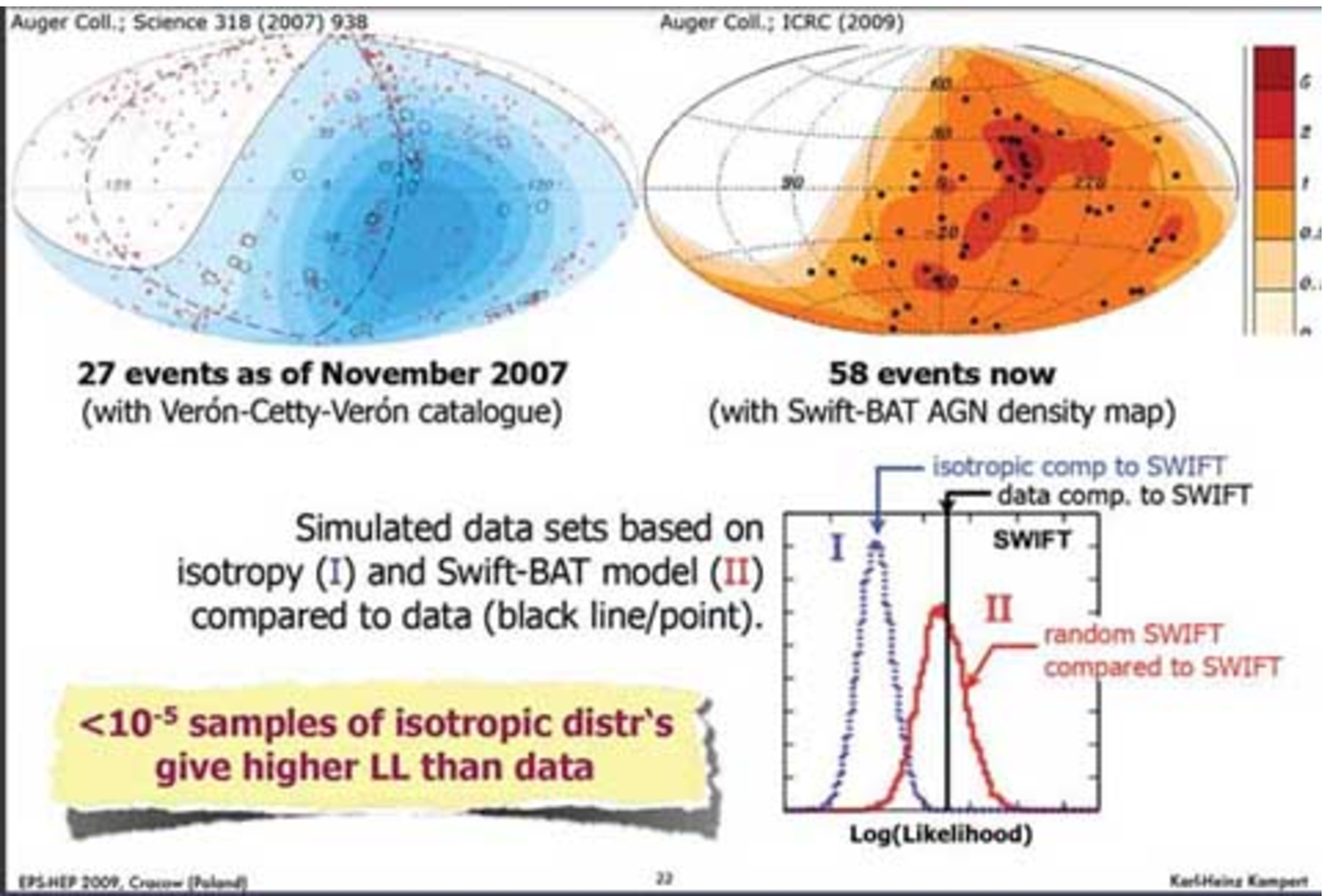}
\includegraphics[width=100pt, angle=0]{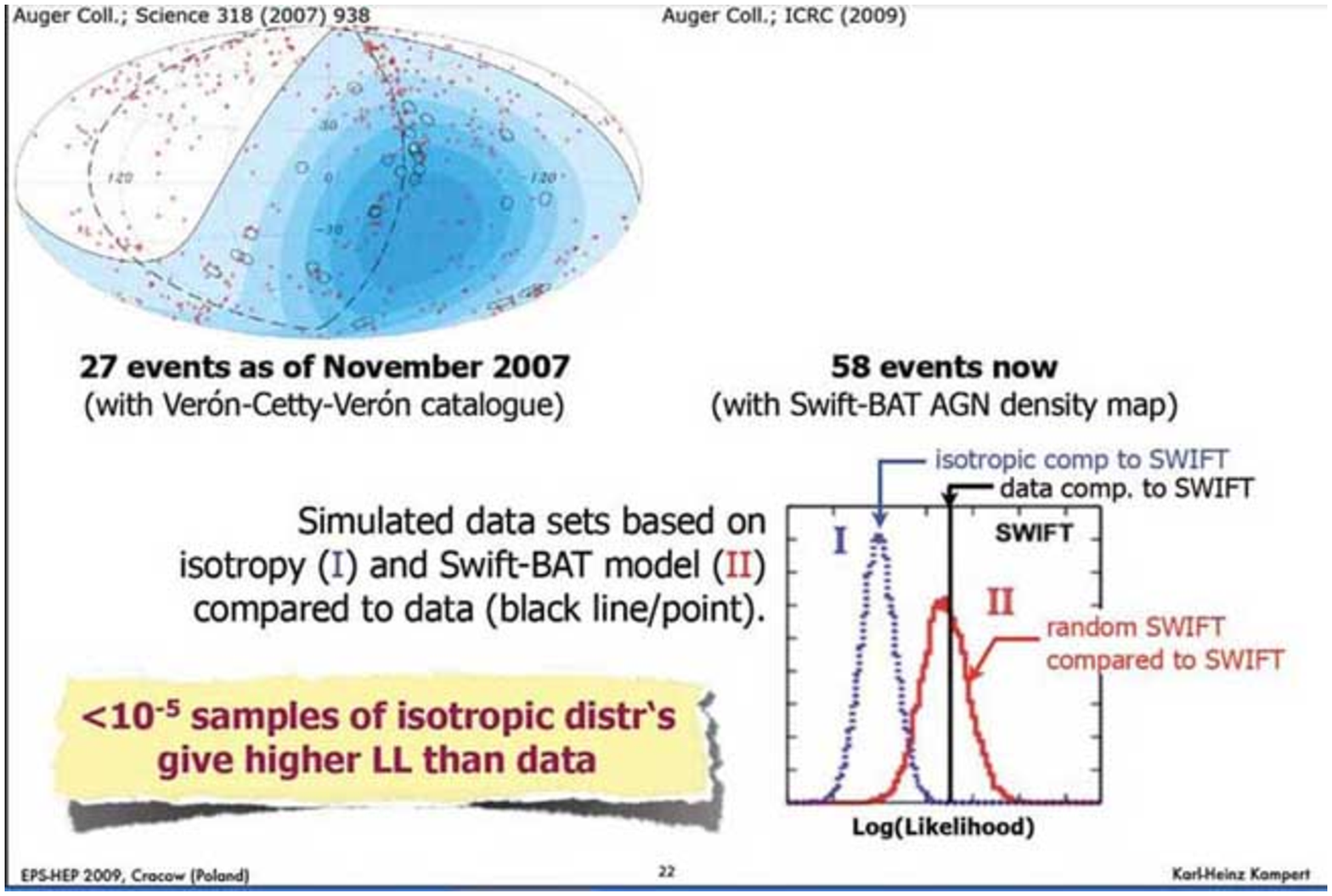}}
\caption{Left: When this paper has been submitted the source of the map \cite{Hep09} was showing the slide on the left; for unknown reason, this slide now faded away: indeed on the right side the same 
slide n.22 as it is found in this day on the web. To avoid any  miss-understanding, an identical UHECR (ring events) map (coincident with the above one), is up to day (26th November 2009) public on web in the related
site: "http://agenda.infn.it/conferenceOtherViews.py?view=standard$\&$amp;confId=1369", Contribute.21.\cite{Scineghe09}, as discussed and shown  in a more recent paper ,\cite{Fargion2009c}.
}
\end{figure*}

\section*{Conclusion: Neutrinos Shadows}
The common expected photo-pion secondary (by proton-photon GZK)  tau Neutrino  are leading to tau air-shower at EeV detectable by a fluorescence tails mostly at far edges ($15-25$ km) distances from the telescopes. Also by showering and skimming on the Surface Detectors \cite{Fargion1999}\cite{Feng2002}. Such EeV GZK Tau-Air-showers would usually rise at smaller inclination (about $2.5^o$ ) because Earth opacity. Such GZK EeV events are contained at AUGER or TA view, at far distances and within a  narrow azimuth angle (about $20^o-30^o$). Tau Air-shower detection for GZK neutrinos has been proposed since a decade up to now days \cite{Fargion1999}, \cite{Fargion2000},\cite{Bertou2002},\cite{Feng2002},\cite{Auger08}. Within our present view the ruling secondaries will be tens-hundred PeV neutrinos observable very near the AUGER or Telescope Array eyes, appearing at much wider azimuth angle (about $120^o-150^o$). Therefore future UHE GZK neutrino astronomy map  may reflect both a spread UHECR lightest nuclei tail and a correlated EeV neutrino born near the source. A neutrino spectroscopy via GZK in the sky may overlap  a prompt point source origination. This tau neutrino astronomy  may soon show by its spectra (tens PeVs- versus EeV), their parent (shadows) UHECR nature and composition. Unfortunately ICECUBE muon neutrino telescope at tens PeV may play a minor role: PeV muons still suffer of atmospheric noises. Tens PeVs and Glashow resonant neutrino events offer a limited rate in ICECUBE and a negligible directionality (both of tau hadronic or electromagnetic showers) in ice and water. Tau Airshower at tens-hundred PeV at near AUGER-HEAT and TA sky are, in my opinion, at present, a better tool to reveal this Lightest nuclei Neutrino trace. Correlated UHECR secondary fragment tails at 1-3 tens EeV, Tau at tens PeV may soon probe the lightest nuclei model \cite{Fargion2009},\cite{Fargion2009b}.

\bibliographystyle{elsarticle-num}
\bibliography{refs}

\end{document}